\documentclass[prl,twocolumn,showpacs,amsmath,amssymb,superscriptaddress]{revtex4}
\usepackage{pifont,amssymb,epsf,graphicx}
\begin{document}
\title{Zero Energy Bound States on Nano Atomic Line Defect in Iron-based High Temperature Superconductors}

\author{Degang Zhang}
\affiliation{College of Physics and Electronic Engineering, Sichuan Normal University,
Chengdu 610101, China}
\affiliation{Institute of Solid State Physics, Sichuan Normal
University, Chengdu 610101, China}
\affiliation{Texas Center for Superconductivity and Department
of Physics, University of Houston, Houston, Texas 77204, USA}

\begin{abstract}

Motivated by recent scanning tunneling microscopy experiments on Fe atomic line defect in iron-based high temperature superconductors,
we explore the origin of the zero energy bound states near the endpoints of the line defect by employing the two-orbit
four-band tight binding model. With increasing the strength of the Rashba spin-orbit coupling along the line defect,
the zero energy resonance peaks move simultaneously forward to negative energy for $s_{+-}$ pairing symmetry,
but split for $s_{++}$ pairing symmetry. The superconducting order parameter correction due to As(Te, Se) atoms missing
does not shift the zero energy resonance peaks. Such the zero energy bound states are induced by the weak magnetic order
rather than the strong Rashba spin-orbit coupling on Fe atomic line defect.

\end{abstract}


\maketitle

\section{\bf 1. Introduction}

Since the discovery of cuprates in 1986 [1], high temperature superconductivity
has been a focus of both theoretical and experimental investigations in condensed matter physics.
After twenty-two years, another family of high temperature superconductors, i.e. iron-based superconductors,
was also found in 2008 [2]. Such the superconductors with high transition temperatures usually possess
the layered crystal structures  consisting of the conducting planes, e.g. the CuO$_2$ planes
in the cuprates and the Fe-As(Te, Se) layers in the iron-based superconductors.
It is known that the ligand O or As(Te, Se) atoms not only affect heavily the energy band structures,
but also play a crucial role in forming the superconducting pairing symmetries.
The O atoms on the Cu-Cu bonds lead to the d-wave order parameter in the cuprates
while the As(Te, Se) atoms above and below the center of each face of the Fe
square lattice induce the $s_{+-}$ or $s_{++}$ pairing symmetry due to the predominant spin fluctuations
or the orbital fluctuations in the iron-based superconductors.
These superconducting pairing symmetries can be identified by a nonmagnetic impurity in the conducting planes,
which produces a zero energy [3-9] or in-gap bound states or no resonance peaks
in the local density of states (LDOS) [10,11].

\begin{figure}
\rotatebox[origin=c]{0}{\includegraphics[angle=0,
          height=2.0in]{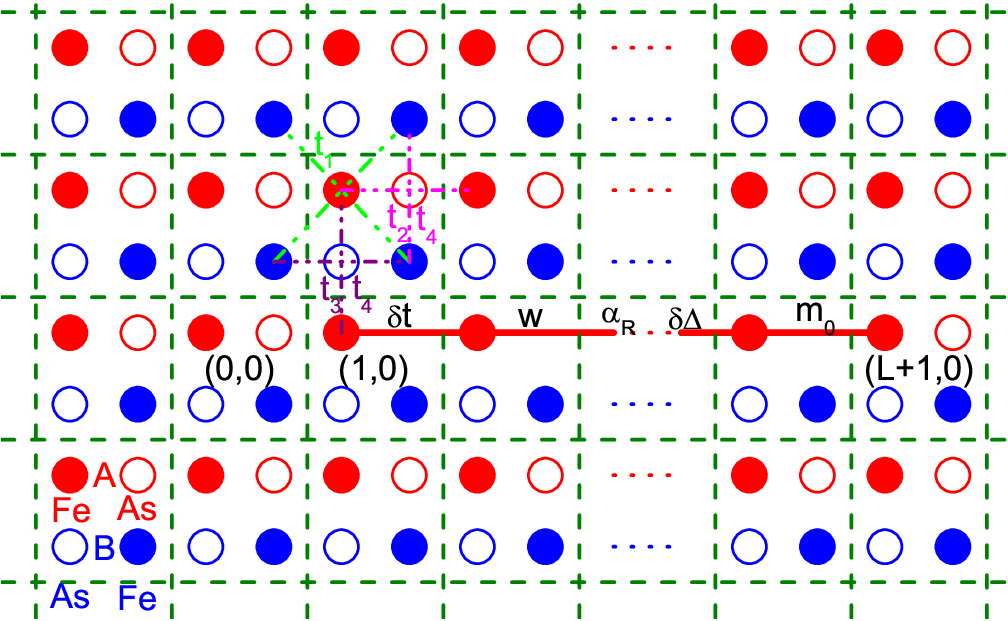}}
\caption {(Color online) Schematic of the atomic line defect produced by L
As (Te, Se) vacancies in the Fe-As (Te, Se) layer with each unit cell containing two
Fe (A and B) and two As (Te, Se) (A and B) ions. The As (Te, Se) ions A and B are
located just above and below the center of each face of the Fe
square lattice, respectively. Here,
$t_1$ is the nearest neighboring
hopping between the same orbitals $d_{xz}$ or $d_{yz}$, $t_2$ and
$t_3$ are the next nearest neighboring hoppings between the same
orbitals mediated by the As (Te, Se) ions B and A, respectively, $t_4$ is the
next nearest neighboring hopping between the different orbitals.
$\delta t$ ($W=-t_4$) and $\delta\Delta$ are the local hopping correction
between the same (different) orbitals and the superconducting order parameter correction
due to the As (Te, Se) vacancies, respectively.
$\alpha_R$ and $M_0$ are the Rashba spin-orbit coupling and the weak magnetic moment
on the atomic line defect.}
\end{figure}

In order to understand the mechanics of high temperature superconductivity,
we first find out the role of the ligands in the high temperature superconductors.
In Ref. [11], a single As vacancy  on the surface of optimally electron-doped BaFe$_{2-x}$Co$_x$As$_2$
was investigated by performing scanning tunneling microscopy (STM) experiments.
A pair of in-gap resonance peaks on the As vacancy was observed.
Such the in-gap bound states can be explained successfully by the two-orbit four-band
tight binding model, which takes the asymmetric effect of up and down ligand
As (Te, Se) atoms on the surface Fe-As (Te, Se) layer into account [10,12].
Recently, a nano Fe atomic line defect (ALD) (see Fig. 1), which is produced by missing a line of Te/Se atoms in
one-unit-cell-thick FeTe$_0.5$Se$_0.5$ films grown on SrTiO3(001) substrates, was also studied by using STM [13].
The robust zero energy resonance peaks (ZERPs) in the $dI/dV$ curves show up near the endpoints of the Fe ALD,
which is approximately 6 (3) nm long (short) with 15 (8) Te/Se atoms missing, but disappear
at the middle part of the long Fe ALD.
The ZERPs are produced due to the strong Rashba spin-orbit coupling (SOC) along the Fe ALD with the inversion symmetry breaking [14].
However, it is hard to induce the strong Rashba SOC along the nano Fe ALD. In addition,
the authors in Ref. [13] do not rule out that the Fe ALD becomes magnetic as the missing Te/Se atoms in the top
sublayer remove half of the spin-orbit coupled Fe-chalcogen bonds.
In this work, we present another explanation of the ZERPs, which are
produced by the weak magnetic order on the Fe ALD. Such a mechanism perfectly fits the STM observations [13].
Here we employ the two-orbit four-band tight binding model mentioned above in order to investigate
the influence of the nano Fe ALD on the LDOS [10].
The empirical energy band model can exhibit excellently the energy band structure of the iron-based superconductors
and its evolution with electron or hole doping measured by ARPES experiments [15-24]
and explain successfully a series of STM experiments
in iron-based superconductors, e.g. in-gap impurity bound states [10,25], the negative energy resonance peak
in the vortex core [26,27], the $90^o$ domain walls and anti-phase domain walls[28-31],
the zero-energy bound state induced by the interstitial excess Fe ions[32-34], etc.,
and especially repeated the phase diagram observed by
nuclear magnetic resonance and neutron scattering experiments [35-37].
Very recently, this tight binding model was also used to study the competition among the superconducting order,
the magnetic order, and the kinetic energy  in europium-based iron pnictides [38].

\begin{figure}
\rotatebox[origin=c]{0}{\includegraphics[angle=0,
           height=1.7in]{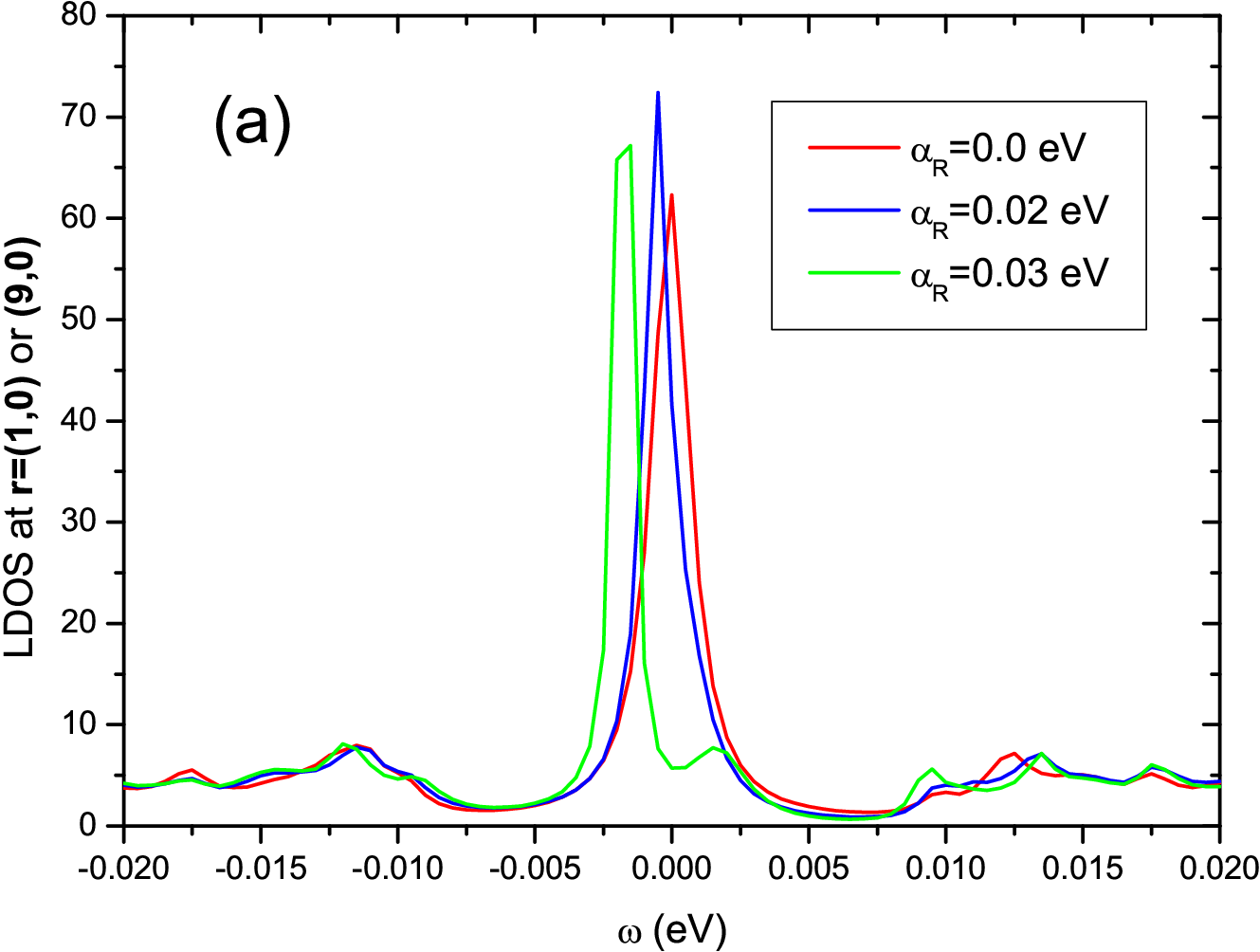}}
\rotatebox[origin=c]{0}{\includegraphics[angle=0,
           height=1.7in]{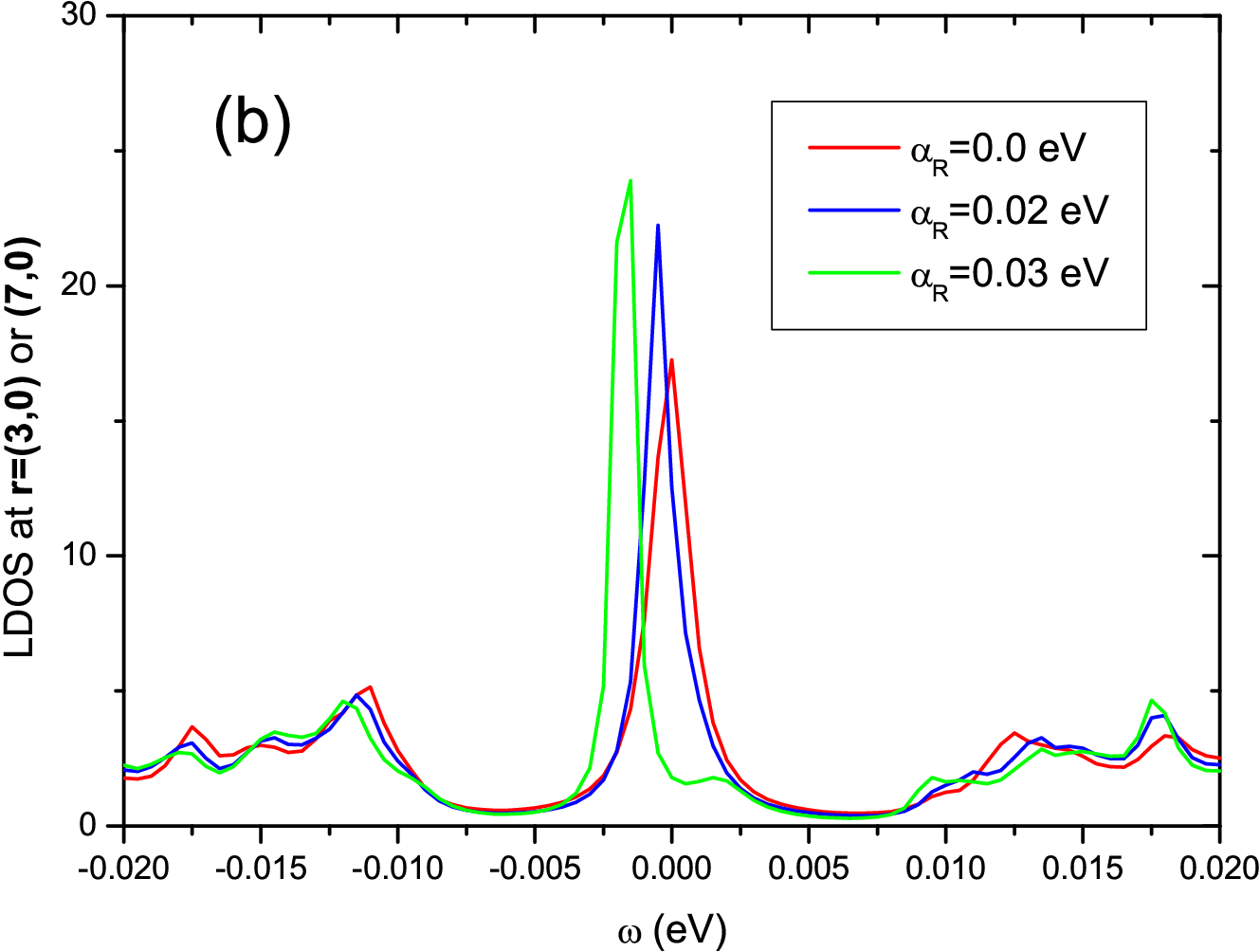}}
\rotatebox[origin=c]{0}{\includegraphics[angle=0,
           height=1.7in]{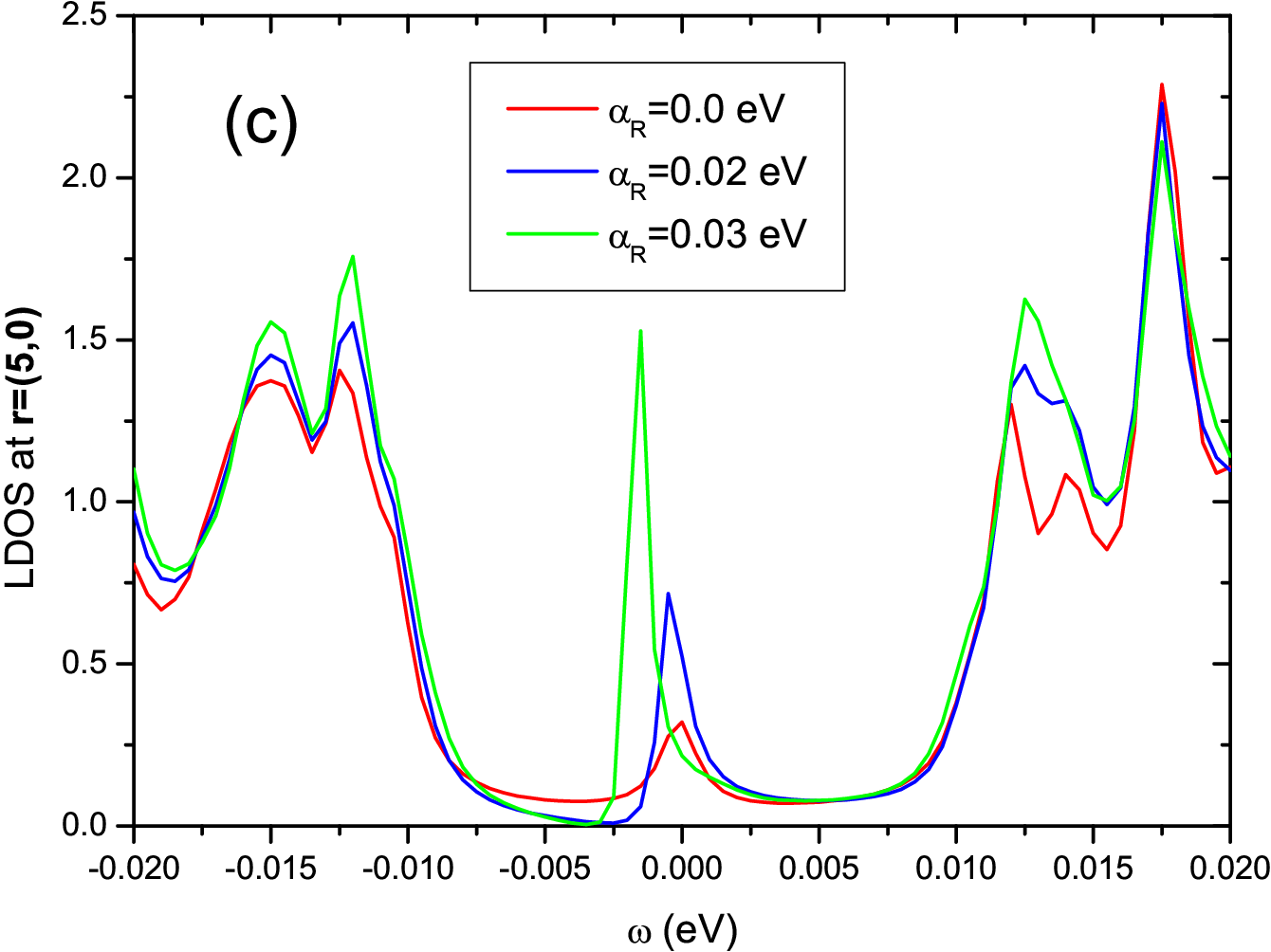}}
\caption {(Color online) The LDOS on the different
Fe sites of the atomic line defect as a
function of the bias voltage $\omega$ under different $\alpha_R$
at optimal electron doping ($15\%$)
for $s_{+-}$ pairing symmetry $
\Delta_{uv{\rm \bf k}}=\frac{1}{2}\Delta_0(\cos k_x+\cos k_y)$.
Here, $\Delta_0=0.018$ eV is the large superconducting energy gap measured by
STM experiments, $\delta t=0.4$ eV, $\delta\Delta=0.0$ eV, and $M_0=0.02$ eV. }
\end{figure}

\section{\bf 2. Model of ALD}

The Hamiltonian describing the Fe ALD in iron-based superconductors can be written as
$$H=H_0+H_{\rm BCS}+H_{ALD}, \eqno{(1)}$$
where $H_0$ and $H_{\rm BCS}$ are the two-orbit four-band tight binding model
and the mean field BCS pairing Hamiltonian in the Fe-Fe plane, respectively [10],
$H_{ALD}$ is the Hamiltonian induced by the As (Te, Se) vacancies on a line, which has general form
$$\begin{array}{rcl}
H_{ALD}&= &\sum_{j=1}^{L}\{\delta t\sum_{\alpha\sigma}[c^\dagger_{A\alpha(j,0)\sigma}
c_{A\alpha(j+1,0)\sigma} \\
&&+c^\dagger_{B\alpha(j,0)\sigma}
c_{B\alpha(j,1)\sigma}+{\rm h.c.}] \\
&&+W\sum_{\alpha\sigma}[c^\dagger_{A\alpha(j,0)\sigma}c_{A1-\alpha(j+1,0)\sigma}\\
&&+c^\dagger_{B\alpha(j,0)\sigma}c_{B1-\alpha(j,1)\sigma}+{\rm h.c.}]\\
&&+\delta\Delta\sum_{\alpha}[c^\dagger_{A\alpha(j,0)\uparrow}
c^\dagger_{A\alpha(j+1,0)\downarrow} \\
&&+c^\dagger_{A\alpha(j,0)\downarrow}
c^\dagger_{A\alpha(j+1,0)\uparrow}\\
&&+c^\dagger_{B\alpha(j,0)\uparrow}
c^\dagger_{B\alpha(j,1)\downarrow} \\
&&+c^\dagger_{B\alpha(j,0)\downarrow}
c^\dagger_{B\alpha(j,1)\uparrow}+{\rm h.c.}]\\
&&+i\alpha_R\sum_{\alpha\sigma\sigma^\prime}[c^\dagger_{A\alpha(j,0)\sigma}{s}_z^{\sigma\sigma^\prime}
c_{A\alpha(j+1,0)\sigma^\prime}\\
&&-c^\dagger_{A\alpha(j+1,0)\sigma}{s}_z^{\sigma\sigma^\prime}
c_{A\alpha(j,0)\sigma^\prime}]\}\\
&&+M_0\sum_{j=1}^{L+1}\sum_{\alpha}[c^\dagger_{A\alpha(j,0)\uparrow}
c_{A\alpha(j,0)\uparrow}\\
&&-c^\dagger_{A\alpha(j,0)\downarrow}
c_{A\alpha(j,0)\downarrow}].\\
\end{array} \eqno{(2)}$$
Here, $\alpha=0$ and $1$  represent  the degenerate orbitals $d_{xz}$ and $d_{yz}$,
respectively,  $s_z$ is the Pauli matrix along the z direction,
$c^+_{A(B),\alpha,(i,j),\sigma }$
($c_{A(B),\alpha,(i,j),\sigma}$) creates (destroys) an $\alpha$
electron with spin $\sigma$ (=$\uparrow$ or $\downarrow$) in the unit cell
$(i, j)$ of the Fe sublattice A (B).
$\delta t$ ($W$) is the local hopping correction between the same (different) orbitals
due to the As (Te, Se) vacancies. Because the As (Te, Se) vacancies cannot mix
$d_{xz}$ orbital and $d_{yz}$ orbital, we always have $W=-t_4$. $\delta\Delta$ is the
superconducting order parameter correction, $\alpha_R$ is the Rashba SOC
induced by the inversion symmetry breaking along the Fe ALD,
and $M_0$ is the magnetic moment produced by the asymmetric environments around the Fe ALD.

\begin{figure}
\rotatebox[origin=c]{0}{\includegraphics[angle=0,
           height=1.7in]{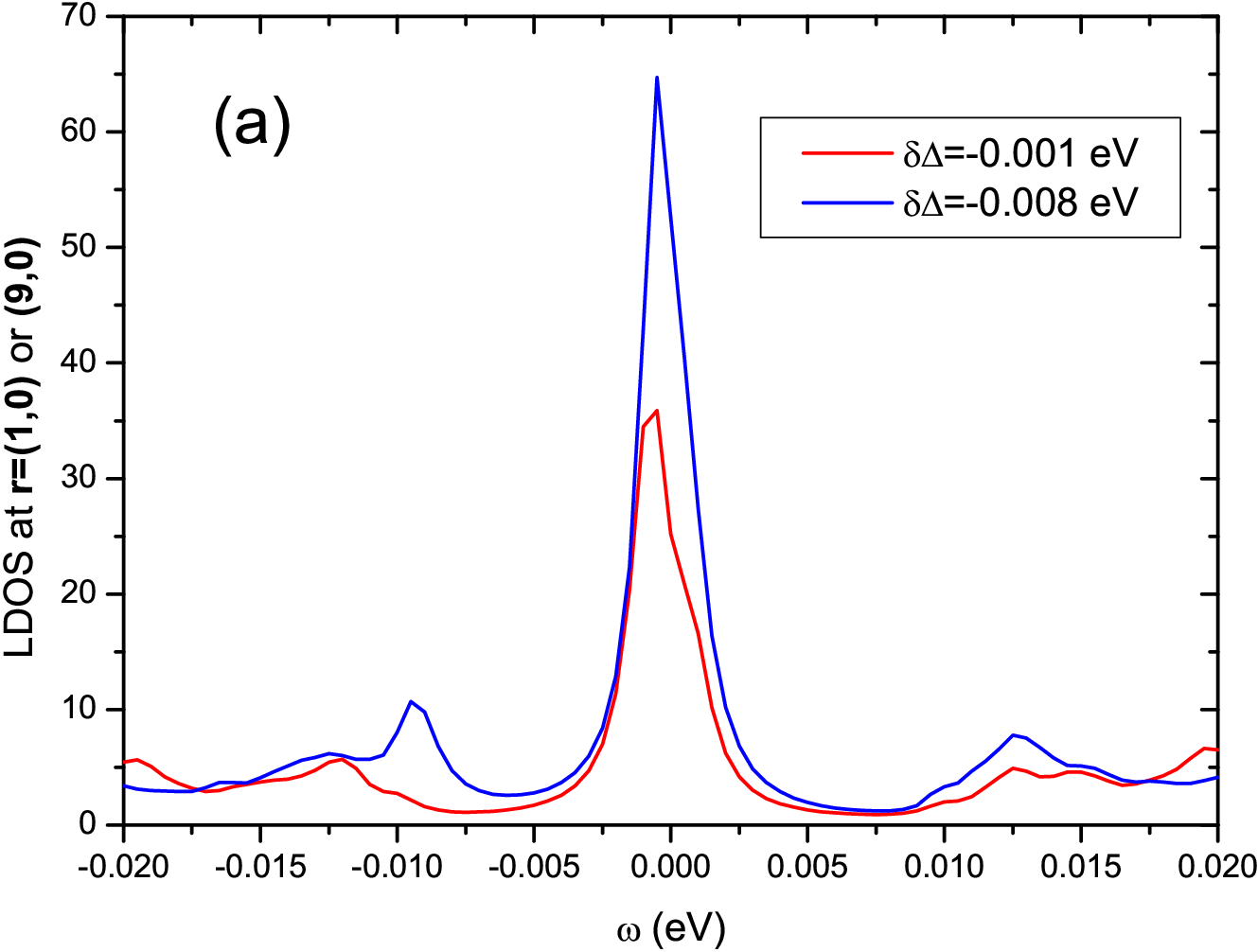}}
\rotatebox[origin=c]{0}{\includegraphics[angle=0,
           height=1.7in]{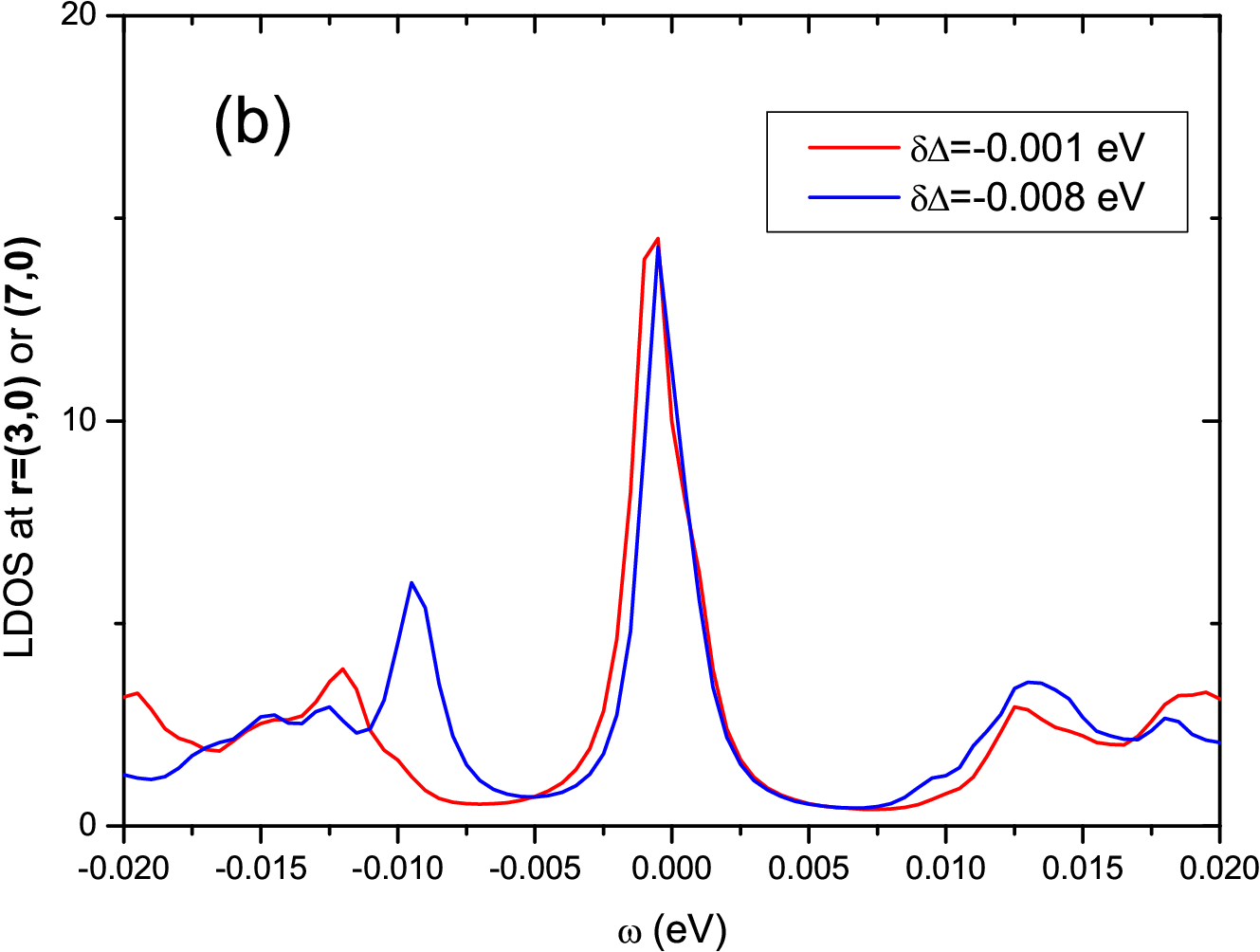}}
\rotatebox[origin=c]{0}{\includegraphics[angle=0,
           height=1.7in]{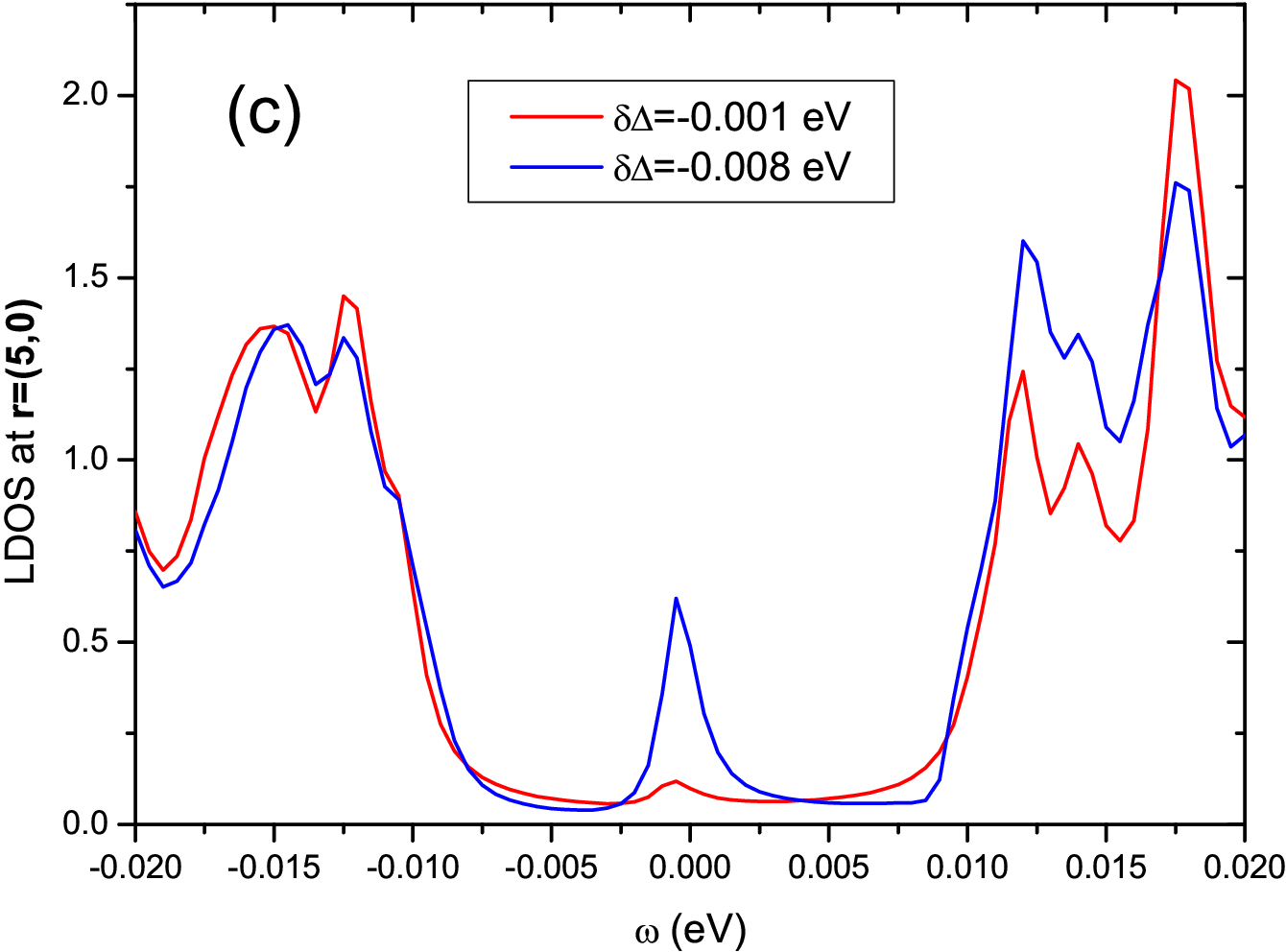}}
\caption {(Color online) The LDOS on the different
Fe sites of the atomic line defect as a
function of the bias voltage $\omega$ under different $\delta\Delta$
at optimal electron doping ($15\%$)
for $s_{+-}$ pairing symmetry $
\Delta_{uv{\rm \bf k}}=\frac{1}{2}\Delta_0(\cos k_x+\cos k_y)$.
Here, $\Delta_0=18.0$ meV is the large superconducting energy gap measured by
STM experiments, $\delta t=0.4$ eV, $\alpha_R=0.008$ eV, and $M_0=0.02$ eV.}
\end{figure}

After introducing first the Fourier transformations
${c}_{A(B)\alpha(i,j)\sigma}=\frac{1}{\sqrt{N}}\sum_{\bf
k}{c}_{A(B)\alpha{\bf k}\sigma}e^{i(k_x x_i+k_y y_j)}$ with $N$ the
number of unit cells and the canonical transformations
for ${c}_{A,\alpha,{\bf k},\sigma}$ and ${c}_{B,\alpha,{\bf k},\sigma}$,
and then taking the Bogoliubov transformations for new
fermion operators, we can also solve analytically the Hamiltonian $H$ for the
Fe ALD in iron-based superconductors by using the T-matrix approach [10,12,33].
The analytic formulas for the Green's functions in momentum space
have been derived. The LDOS on the Fe ALD at different bias voltages can be obtained
through the Fourier transformation of the Green's functions in momentum space
and $i\omega_n\rightarrow \omega+i\delta$.
In order to compare with the STM experiments [13],
here we have calculated the short Fe ALD (L=8) in a square Fe lattice
with $N=60\times 60$ unit cells, which is enough to ensure the
accuracy of theoretical results. In our calculations, we have used the energy band parameters:
$t_1=-0.5$ eV, $t_2=-0.2$ eV, $t_3=1.0$ eV,
and $t_4=-0.02$ eV, which are same with the previous works
[10,12,27,30,31,33,34,37,38], the chemical potential $\mu=-0.49$ eV corresponding to $15\%$
electron doping, the hopping correction $\delta t=0.4$ eV,
the superconducting energy gap $\Delta_0=18$ meV observed by the STM experiments
on the one-unit-cell-thick FeTe$_0.5$Se$_0.5$ films grown on SrTiO3(001) substrates [13],
and $\delta=0.0008$ eV.

\section{\bf 3. Results and discussion}

We plot the curves of the LDOS on the Fe ALD as a function of the bias voltage
$\omega$ under different $\alpha_R$ at the optimal electron doping
(15$\%$) for the $s_{+-}$ pairing symmetry
$\Delta_{uv{\rm \bf k}}=\frac{1}{2}\Delta_0(\cos k_x+\cos k_y)$
in Figs. 2, where $\Delta_0=0.018$ eV, $\delta t=0.4$ eV, $\delta\Delta=0.0$ eV, and $M_0=0.02$ eV.
Obviously, when $\alpha_R$ is weak, the LDOS at the different sites of the Fe ALD
have a zero energy resonance peak (ZERP). The height of the ZERP rapidly decays
with the distance to the endpoint of the Fe ALD. Such a feature of the ZERP
coincides qualitatively with the STM observations [13].
With increasing $\alpha_R$, the ZERPs move simultaneously forward to the bias voltage.
We note that if $\alpha_R$ exceeds a critical value, the LDOS at the midpoint, i.e. ${\bf r}=(5,0)$,
is negative under some bias voltages, which is unphysical (see Fig. 2c).
Therefore, the Rashba SOC on the Fe ALD is not too strong.

\begin{figure}
\rotatebox[origin=c]{0}{\includegraphics[angle=0,
           height=1.7in]{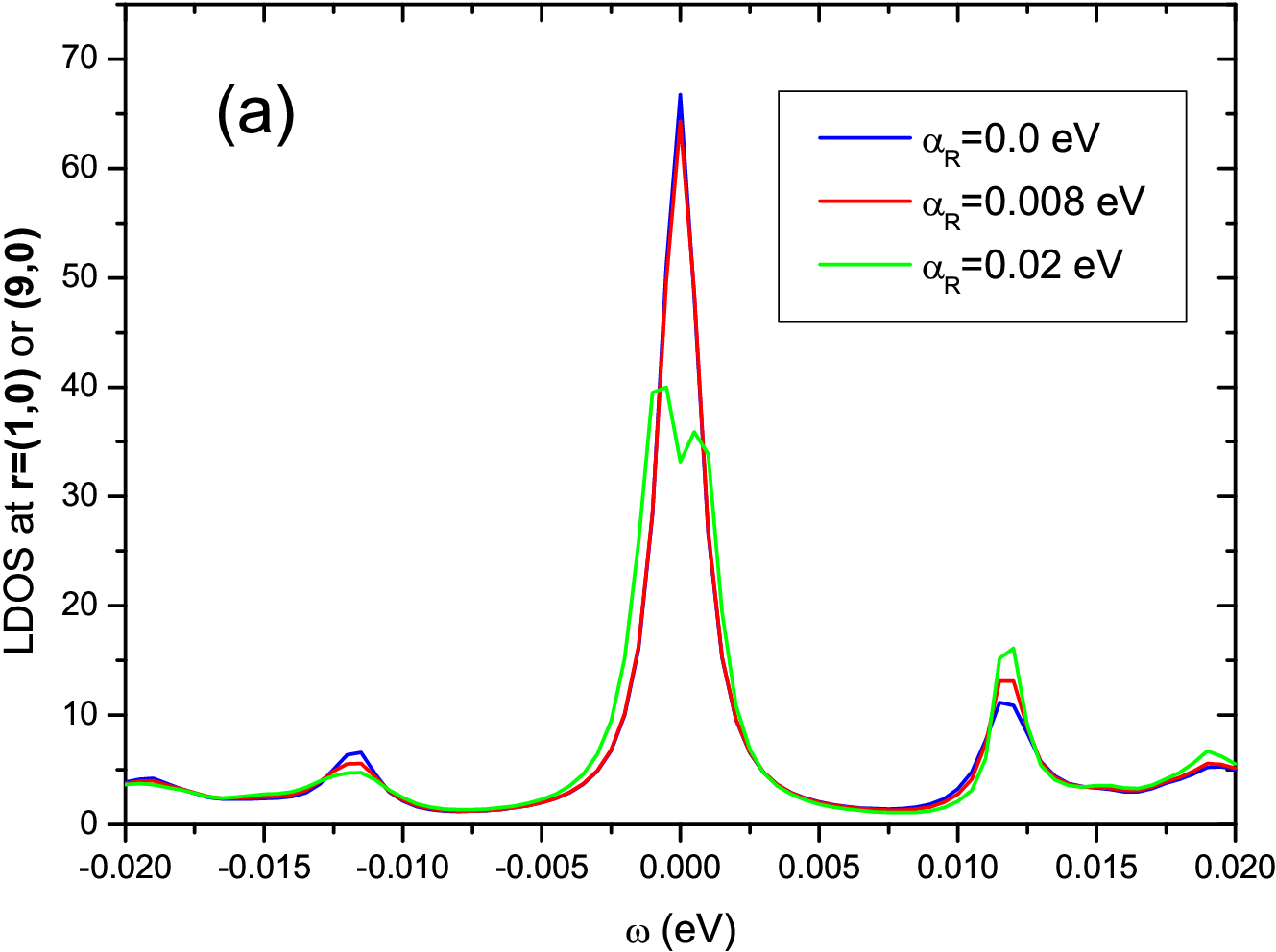}}
\rotatebox[origin=c]{0}{\includegraphics[angle=0,
           height=1.7in]{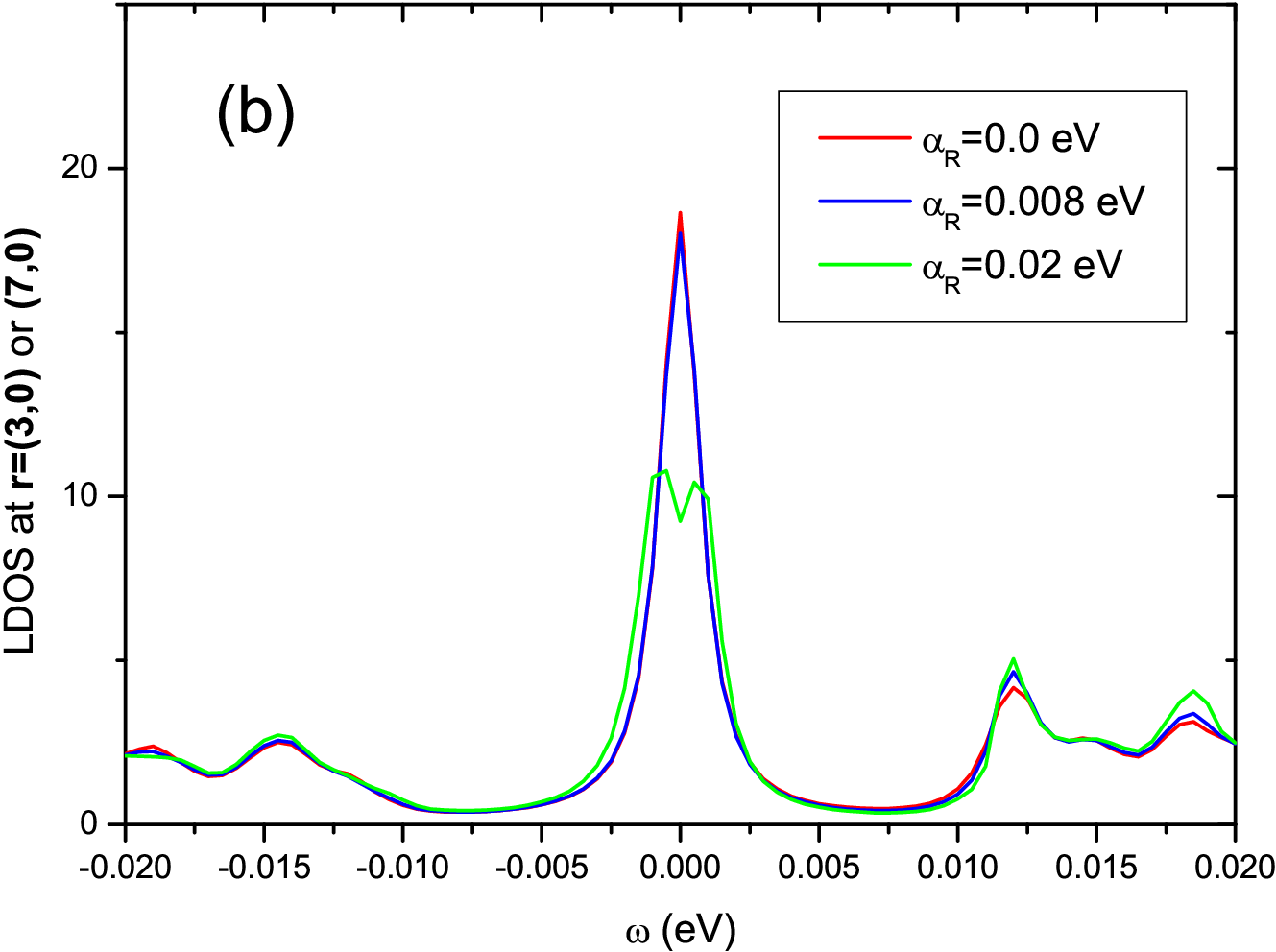}}
\rotatebox[origin=c]{0}{\includegraphics[angle=0,
           height=1.7in]{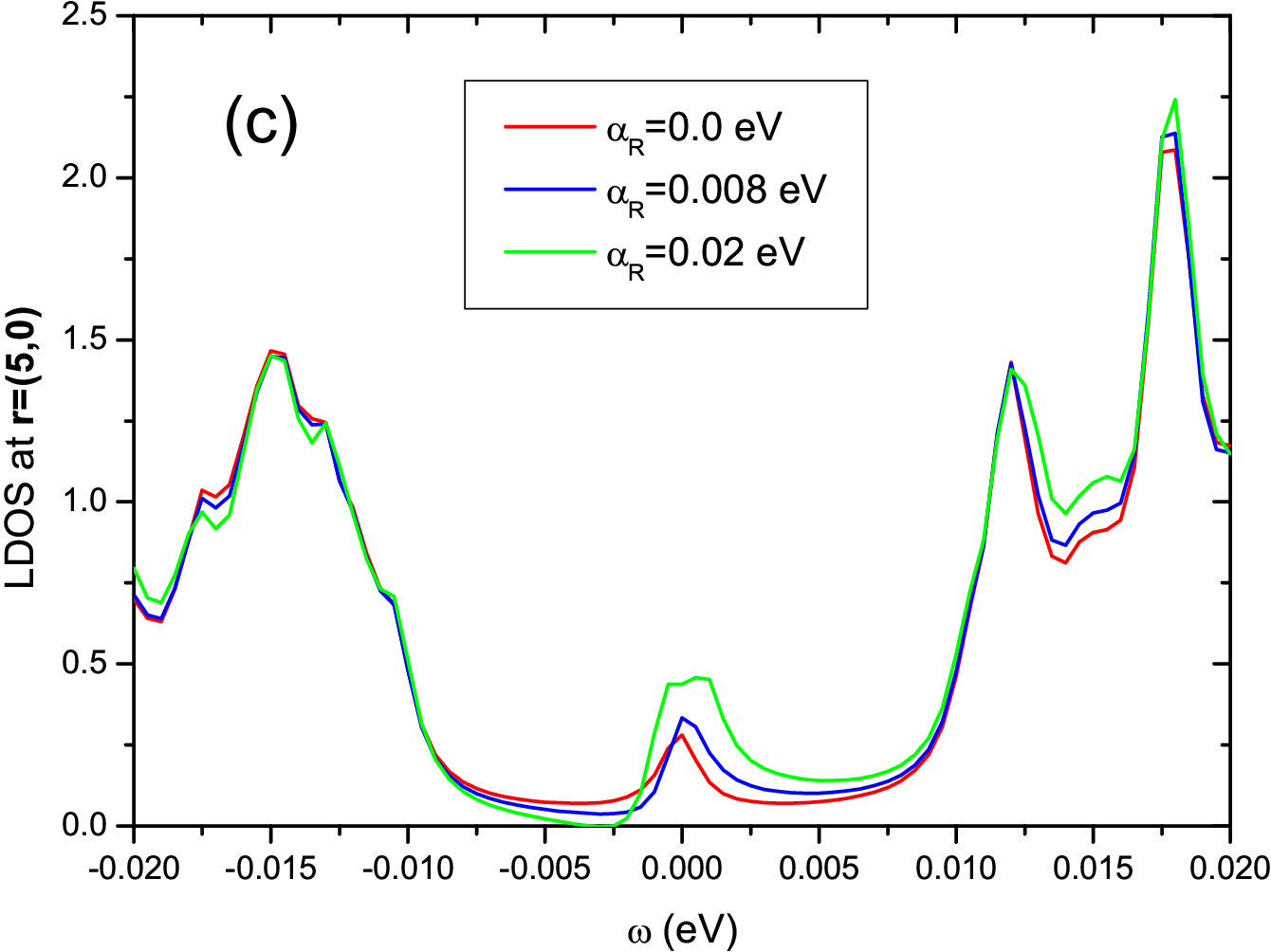}}
\caption {(Color online) The LDOS on the different
Fe sites of the atomic line defect as a
function of the bias voltage $\omega$ under different $\alpha_R$
at optimal electron doping ($15\%$)
for $s_{++}$ pairing symmetry $
\Delta_{uv{\rm \bf k}}=\frac{1}{2}\Delta_0|(\cos k_x+\cos k_y)|$.
Here, $\Delta_0=0.018$ eV is the large superconducting energy gap measured by
STM experiments, $\delta t=0.4$ eV, $\delta\Delta=0.0$ eV, and $M_0=0.025$ eV.}
\end{figure}

Fig. 3 shows the LDOS on the Fe ALD as a function of the bias voltage
$\omega$ under different $\delta\Delta$ at the optimal electron doping
(15$\%$) for the $s_{+-}$ pairing symmetry when $\alpha_R=0.008$ eV.
With increasing $|\delta\Delta|$,
the location of the ZERP keeps unchanged. However, the ZERP becomes higher
at the endpoint ${\bf r}=(1,0)$ or ${\bf r}=(9,0)$ and the midpoint ${\bf r}=(5,0)$.
Meanwhile, the locations of the superconducting coherence peaks (SCP) have a large shift
forward to zero energy near the endpoints of the Fe ALD and seem not to change at the midpoint.

In Fig. 4, we depict the LDOS on the Fe ALD as a function of the bias voltage
$\omega$ under different $\alpha_R$ at the optimal electron doping
(15$\%$) for the $s_{++}$ pairing symmetry
$\Delta_{uv{\rm \bf k}}=\frac{1}{2}\Delta_0|(\cos k_x+\cos k_y)|$
when $\Delta_0=0.018$ eV, $\delta t=0.4$ eV, $\delta\Delta=0.0$ eV, and $M_0=0.025$ eV.
If $\alpha_R$ is small, the LDOS at the different sites also
possess a ZERP. The height of the ZERP also decreases rapidly
with the distance to the endpoint of the Fe ALD, similar to the case of the $s_{+-}$ pairing symmetry
in Fig. 2. When $\alpha_R$ becomes larger, the ZERPs simultaneously split, but the locations of the SCP are not shifted.
Obviously, $\alpha_R$ also has a critical value, at which the LDOS at the midpoint is negative under
some bias voltages.

\begin{figure}
\rotatebox[origin=c]{0}{\includegraphics[angle=0,
           height=1.7in]{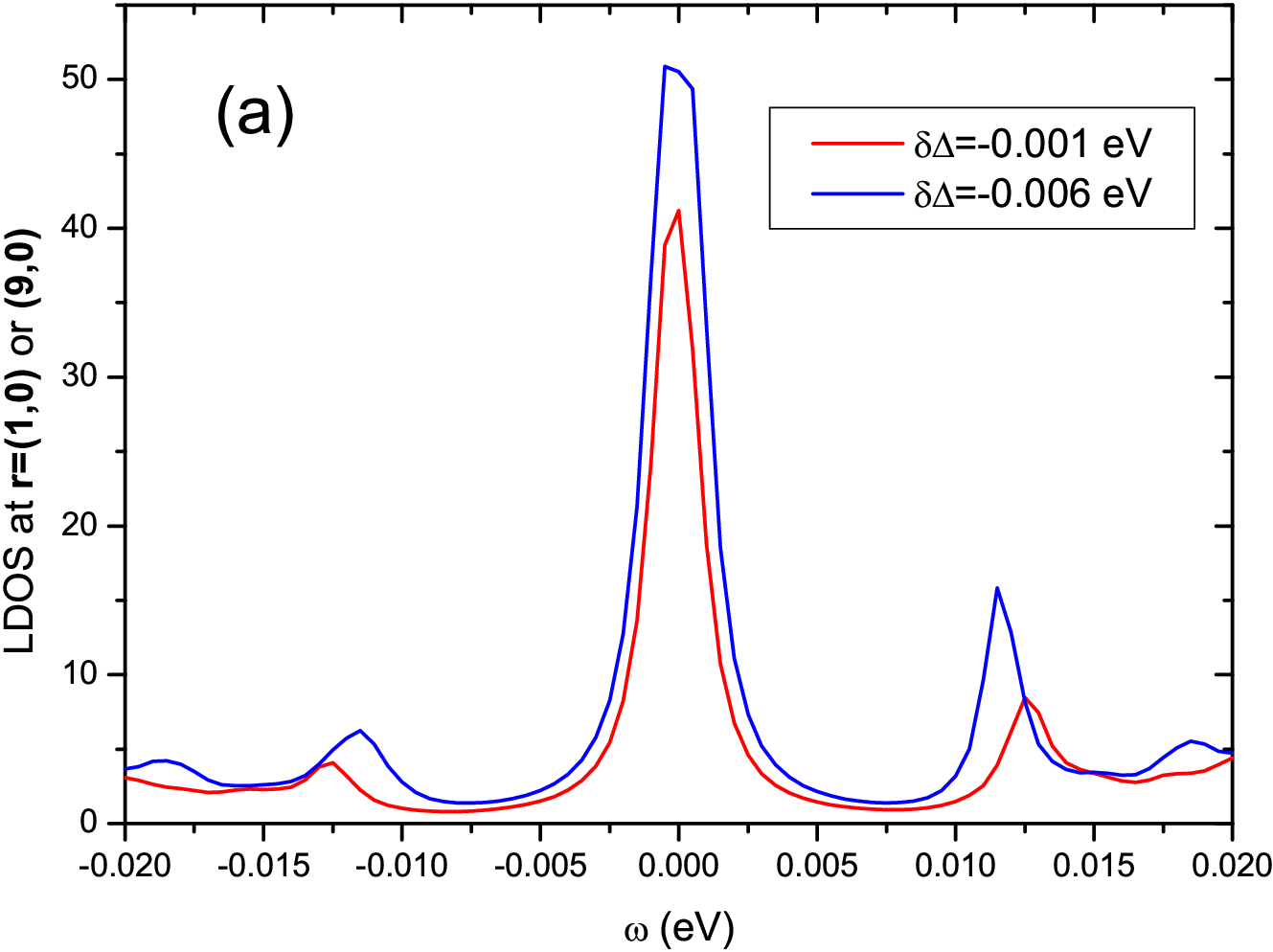}}
\rotatebox[origin=c]{0}{\includegraphics[angle=0,
           height=1.7in]{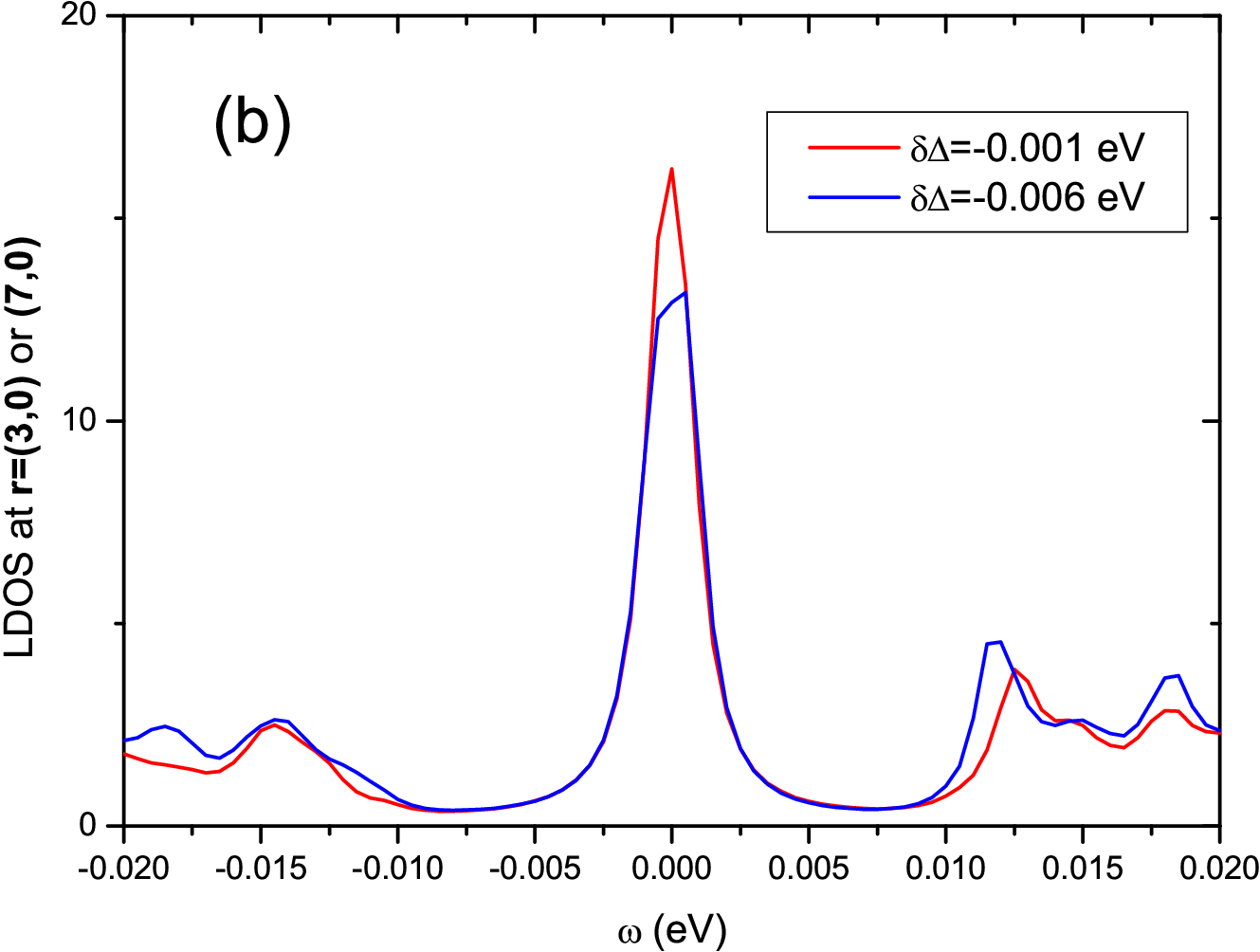}}
\rotatebox[origin=c]{0}{\includegraphics[angle=0,
           height=1.7in]{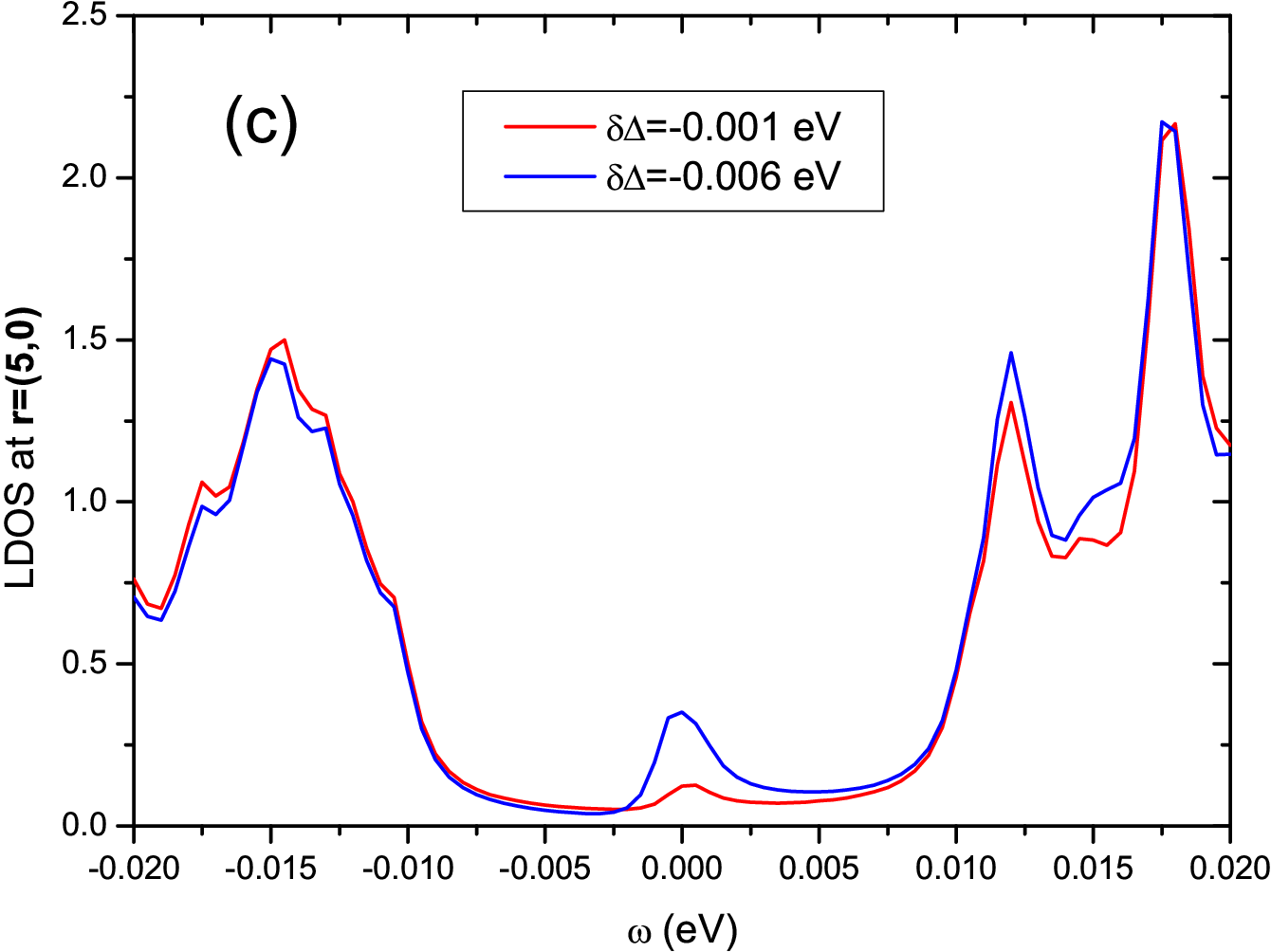}}
\caption {(Color online) The LDOS on the different
Fe sites of the atomic line defect as a
function of the bias voltage $\omega$ under different $\delta\Delta$
at optimal electron doping ($15\%$)
for $s_{++}$ pairing symmetry $
\Delta_{uv{\rm \bf k}}=\frac{1}{2}\Delta_0|(\cos k_x+\cos k_y)|$.
Here, $\Delta_0=18.0$ meV is the large superconducting energy gap measured by
STM experiments, $\delta t=0.4$ eV, $\alpha_R=0.008$ eV, and $M_0=0.025$ eV.}
\end{figure}

Fig. 5 exhibits the LDOS on the Fe ALD as a function of the bias voltage
$\omega$ under different $\delta\Delta$ at the optimal electron doping
(15$\%$) for the $s_{++}$ pairing symmetry. With increasing $|\delta\Delta|$,
the ZERP does not move. The locations of the SCP also have a large shift near the
endpoints of the Fe ALD and seem not to change at the midpoint,
consistent with the case of the $s_{+-}$ pairing symmetry
in Fig. 3.

In summary, we have explored the impact of As (Te, Se) atoms missing on the electronic states
at the Fe ALD in iron-based superconductors. When the Rashba SOC $\alpha_R$ and
the magnetic order $M_0$ are weak, a ZERP appears apparently near the endpoints of the Fe ALD
for both $s_{+-}$ and $s_{++}$ pairing symmetries. The height of the ZERP decays rapidly with
increasing the distance to the endpoint. We note that the ZERP vanishes at the middle part of
a long Fe ALD. Such the ZERPs are consistent qualitatively with the STM experiments [13].
Here we have presented another origin of the ZERPs, which is due to the weak magnetic order rather than
the strong Rashba SOC on the Fe ALD. Such a weak magnetic order
could be detected by neutron scattering experiments.

\section{\bf Acknowledgements}

This work was supported by the Sichuan Normal University, the "Thousand
Talents Program" of Sichuan Province, China, the Texas Center
for Superconductivity at the University of Houston,
and by the Robert A. Welch Foundation under grant No. E-1146.

\end{document}